\documentclass{article}%
\usepackage{amsmath}%
\usepackage{amsfonts}%
\usepackage{amssymb}%
\usepackage{psfig,graphicx,amssymb}
\usepackage{graphicx}
\providecommand{\U}[1]{\protect\rule{.1in}{.1in}}

\begin{document}

\title{Gluing Feynman diagrams in NDIM:\\ Insights into the three-point vertex}
\author{A. T. Suzuki$^{1,a}$, A. G. M. Schmidt$^{2,b}$, J. D. Bolzan$^{1,c}$, \\
\\$^{1}$Instituto de F\'{\i}sica Te\'{o}rica\\Universidade Estadual Paulista\\Rua Pamplona 145\\01405-900 - S\~{a}o Paulo, SP \\Brazil\\
\\$^{2}$Departamento de Ci\^encias Exatas, Universidade Federal Fluminense\\Av. dos Trabalhadores, 420\\27255-125 - Volta Redonda, RJ\\Brazil}
\maketitle

\begin{abstract}
Three-point vertex diagram plays a key role in the whole renormalization program of several QFT (quantum field theory) models such as QED, QCD, the Standard Model of eletroweak interactions and so forth. The exact analytic result for the triangle diagram therefore is fundamental.

In this work we calculate in two different ways a two-point two-loop massless Feynman
diagram using what we call a ``gluing'' technique in the context of NDIM (Negative
Dimensional Integration Method). The two-loop diagram in question can be ``glued'' in two different ways and we show that both yield the same result and reproduce the one calculated via NDIM for the complete diagram, which, of course, is equivalent to the exact solution obtained by normal positive dimensional calculation.

Furthermore, in the process we conclude that the usual massless off-shell triangle diagram result does not hold anymore and present a new solution for it with only three hypergeometric functions $F_{4}$.

\vspace{.3cm}


\end{abstract}

\vspace{.3cm}

$^{a}${\footnotesize E-mail: suzuki@ift.unesp.br}

$^{b}${\footnotesize E-mail: agmschmidt@pq.cnpq.br}

$^{c}${\footnotesize E-mail: jbolzan@ift.unesp.br}

\section{Introduction}

Today in quantum field theory, it is necessary to calculate increasingly
complex Feynman diagrams as the theory and the experiments require a higher
accuracy of the scattering amplitudes; a very good example of this being Kinoshita's quest: to calculate $(g-2)$ up to $\alpha^{5}$ order \cite{kinoshita}.
Several techniques have been applied for that purpose --- most of them in the
context of dimensional regularization \cite{dreg1} or analytic regularization
\cite{cicuta} --- and among them we can mention the powerful Mellin-Barnes contour
integration \cite{mb1, mb2, mb3}, the method of Gegenbauer polynomials
\cite{gegenbauer}, the differential equations technique \cite{remiddi} and
others \cite{laporta}. The NDIM developed by Halliday and Ricotta
\cite{halliday} has shown itself as a reliable one when applied to the
calculation of diagrams of one- \cite{glover, esdras}, two- \cite{2loops} and
multi-loops \cite{ivan, achievements}, with scalar and tensorial structures
and in noncovariant gauges \cite{tensor-gauge}. One of the advantages of NDIM is that it allows us
to avoid the often cumbersome parametric integrals, transfering the problem into easier
solving systems of linear equations instead. Another advantage of NDIM is that the exponents of
propagators are taken to be arbitrary integers, so that one can solve the general case for each type of graph.

Despite of having these advantageous features which turn itself an efficient and simple
method, NDIM has some drawbacks pointed out in earlier works:
the large number of systems of equations and the difficulty in dealing with the ever increasing complexity of the hypergeometric type series that results. Related to this difficulty is, for example, the summing up analytically of hypergeometric series with unit argument, such as, $_{p}F_{q}(...|1)
$ that often appears in two-point function calculations. The approach to overcome the first difficulty, namely, to reduce the
growing number of linear systems --- it grows with the number of loops and
legs attached to the diagram at hand --- was presented by Gonzalez and Schmidt
for massless diagrams: they proposed a way to write down the generating
gaussian integral in order to optimize (to minimize!) such number. Then the
whole calculation can be made simpler and faster, and the number of
hypergeometric functions $_{p}F_{q}(...|1)$ left in the final result is also minimum.

A great effort is being conducted in order to study maximally supersymmetric
Yang-Mills theory (MSYM). Several authors \cite{msym} are tackling a rather
difficult task: to calculate certain scattering amplitudes exactly since
Maldacena conjectures that higher loop contributions can be written in terms
of one-loop amplitudes. Tests of this conjecture have been realized from
4-point 2-loops to even more challenging 4-point 5-loops Feynman amplitudes,
involving the so-called dual conformal integrals. The outcome of this program
can be a resummation of the entire perturbative series for a given physical
process. The reason to have powerful methods to tame these integrals is clear.

We present in this paper another way to apply NDIM in Feynman integral calculations, the ``gluing'' approach. In general a Feynman diagram, e.g., the 2-loop master diagram, is represented
by an integral,

\begin{equation}
I=\int\int\frac{d^{D}k\;d^{D}q}{(k^{2})^{i}[(k-p)^{2}]^{j}[(k-q)^{2}%
]^{l}(q^{2})^{m}[(q-p)^{2}]^{n}},
\end{equation}
where $p$ is the external momentum, and $i,j,l,m,n$ are exponents of
propagators, which can be made arbitrary in the whole calculation. One could
rewrite the above integral as,
\begin{equation}
I=\int\frac{d^{D}k}{(k^{2})^{i}[(k-p)^{2}]^{j}}\int\frac{d^{D}q}%
{[(k-q)^{2}]^{l}(q^{2})^{m}[(q-p)^{2}]^{n}},\label{1st_integral}%
\end{equation}
and we readily recognize the integral in momentum $q$ as an off-shell triangle
one, which has a well-known result \cite{boos} that can be written in terms of four Appel
hypergeometric functions of two variables $F_{4}(...|k^{2}/p^{2},(k-p)^{2}/p^{2})$. It
is straightforward to see that the remaining integral is a self-energy one
with shifted exponents of propagators,
\begin{equation}
I=\Gamma\int\frac{d^{D}k}{(k^{2})^{\nu_{1}}[(k-p)^{2}]^{\nu_{2}}%
},\label{2nd_integral}%
\end{equation}
where $\Gamma$ is a factor which depends on $p^{2}$ as well as the exponents
of propagators and dimension $D$. The new exponents $\nu_{1}$ and $\nu_{2}$
also depend of the former ones $i,j,l,m,n$, as well as of dimension and sum
indices of $F_{4}$ functions. However, straightforward application of this does not yield the correct result. Here comes an important point: to carry out the
second integral (\ref{2nd_integral}) one has to perform the integral over the
whole space, for this reason the result of the former one must hold on the
whole range of momentum $k$. The well-known result of the off-shell triangle,
written as a sum of four Appel's hypergeometric functions $F_{4}(...|x,y)$,
\textit{is not valid} for every momentum; these momenta must be such that
$|x|<1$, $|y|<1$ and $|\sqrt{x}|+|\sqrt{y}|<1$. In other words, the series is
defined inside some region of convergence and for this reason the well-known
result of Boos and Davydychev \cite{esdras, boos} can not be used in
(\ref{1st_integral}).

In this paper we use NDIM to solve a massless two-loop self energy diagram in
a different approach as used before \cite{2loops3}, namely, integrating loop
by loop. We separate the diagram into two simpler parts, each one a single
loop diagram itself and then solve these parts ``gluing'' them to obtain the
final result. This technique of dealing with subdiagrams could simplify the
solution of larger diagrams that would lead to difficult systems with a great
number of equations and variables. As far as we know only Bierembaum and
Weinzierl \cite{weinzierl} studied some Feynman diagrams using similar ideas
and Mellin-Barnes method. Kostrykin and Schrader presented a generalized
Kirchoff rule to ``gluing'' quantum graphs and they relied on a complicated star
product \cite{kostrykin}. ``Gluing'' diagrams is not a trivial task in quantum
mechanics nor in quantum field theory. However, we think that if implemented
such method can simplify computations, being very easy and reliable, since in order to calculate a
multi-loop diagram one could, in principle, use only well-known one-loop
integrals. The main objective of the present work is to elaborate this program
within the NDIM context.

There are two manners to cut the two-loop diagram --- here identified as ``flying saucer''
diagram --- in the particular case presented in this work, as will be shown in
Sections $3$ and $4$. The first manner involves a one-loop self-energy diagram
that we call ``sword-fish'', which is very simple to calculate. The second one
uses the one-loop off-shell triangle diagram, which is written in terms of
only three hypergeometric functions $F_{4}$ --- the form suitable for
integrating the second loop ---, a simplified result accomplished by invoking
momentum conservation \cite{circuito}.

The outline of our paper is as follows: in Section $2$ we show the integral
pertaining to the ``flying saucer'' diagram and present the approaches to make
the cutting. In Section $3$ we solve first the ``sword-fish'' diagram, whereas in Section $4$ we do it via the one-loop triangle diagram first and in Section $5$ we present our concluding remarks.

\section{The flying saucer diagram}

We will consider a particular case of the flying saucer diagram, the so called
``side view'', as shown in Fig. $1$. There is another type of this graph, which
we call flying saucer ``front view''. The difference between them is the
exponent of the propagator of the particle with momentum $k$. Clearly the
following gaussian integral is related to this diagram,
\begin{equation}
I(p^{2};D)=\int d^{D}kd^{D}q\exp[-\alpha q^{2}-\beta k^{2}-\gamma
(p-k)^{2}-\omega(k-q)^{2}]\text{,}\label{gaussiana}%
\end{equation}
in other words it generates the Feynman loop integrals, see
eq.($\ref{original}$).

One can integrate it and compare it with its own Taylor expansion,
\begin{equation}
I(p^{2};D)=\overset{}{\underset{}{\overset{\infty}{\underset{i,j,l,m=0}{\sum}%
}(-1)^{i+j+l+m}\frac{\alpha^{i}\beta^{j}\gamma^{l}\omega^{m}}{i!j!l!m!}
\Im(i,j,l,m;D;p^{2})}}\text{,}\label{taylor}%
\end{equation}
where
\begin{equation}
\Im(i,j,l,m;D;p^{2})=\int d^{D}kd^{D}q(q^{2})^{i}(k^{2})^{j}[(p-k)^{2}%
]^{l}[(k-q)^{2}]^{m}\text{,}\label{original}%
\end{equation}
which is our complete negative dimensional integral. We will show all the
necessary steps for the cut cases in more details in the next sections, but
for now it is enough to know that the solution of the complete diagram is,
before the analytic continuation \cite{2loops3},
\begin{align}
& \left.  \Im(i,j,l,m;D;p^{2})=(-\pi)^{D}(p^{2})^{\sigma}\frac{\Gamma
(1+i)\Gamma(1+l)\Gamma(1+m)}{\Gamma(1+\sigma)\Gamma(1-i-\frac{1}{2}%
D)\Gamma(1-m-\frac{1}{2}D)}\right. \nonumber\\
& \left.  \times\frac{\Gamma(1-\sigma-\frac{1}{2}D)\Gamma(1+i+j+m+\frac{1}%
{2}D)\Gamma(1-i-m-D)}{\Gamma(1-l-\frac{1}{2}D)\Gamma(1+i+m+\frac{1}{2}%
D)\Gamma(1+l-\sigma)}\right.  \text{,}\label{solutioncomplete}%
\end{align}

$\bigskip$%

\begin{figure}[h]
\centering
\includegraphics[height=1.6483in,width=4.7899in]{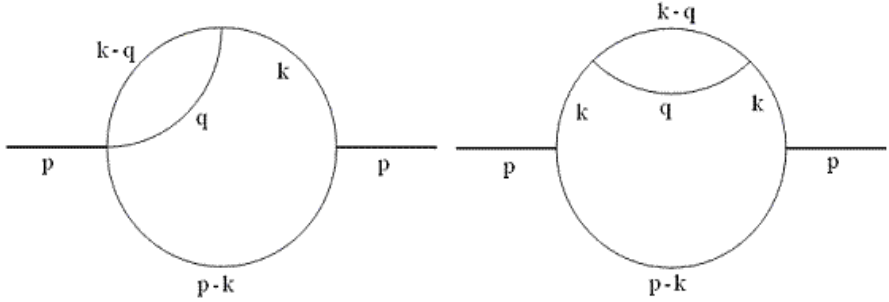}
\caption{Two-point two-loop Feynman diagrams: Fying-saucer side-view and front-view.}
\label{Figure 1:}
\end{figure}

where $\sigma=i+j+l+m+D$.

At this point, our new proposal is to cut the diagram, solve each part separately
and then ``glue'' them together. We expect, of course, that the final result should be the
same as in ($\ref{solutioncomplete}$). In Fig. 2, we show the two
manners to cut this diagram. The first way leaves a one-loop sword-fish graph
plus a vertex, whereas the second way leaves a one-loop triangle and a propagator. In the next sections, we
solve them in details.

\section{The sword-fish diagram}

We begin with the generating integral ($\ref{gaussiana}$), where one can integrate
first in the $q$ momentum,
\begin{equation}
I(p^{2};D)=\int d^{D}k\exp[-\beta k^{2}-\gamma(p-k)^{2}]\int d^{D}%
q\exp[-\alpha q^{2}-\omega(k-q)^{2}]\text{,}\label{q momentum}%
\end{equation}
which represents a one-loop with two massless propagators. Completing the
square, the $q$ integral can be solved easily, giving
\begin{equation}
I_{q}(k^{2};D)=\left(  \frac{\pi}{\alpha+\omega}\right)  ^{\frac{D}{2}}%
\exp\left[  -\frac{\alpha\omega k^{2}}{\alpha+\omega}\right]  \text{.}%
\label{Iq}
\end{equation}

Expanding this result in Taylor series and using the multinomial expansion,
one obtains
\begin{equation}
I_{q}(k^{2};D)=(\pi)^{\frac{D}{2}}\overset{\infty}{\underset{n_{i}=0}{\sum}%
}\frac{(-k^{2})^{n_{1}}(-n_{1}-\frac{1}{2}D)!\alpha^{n_{1}+n_{2}}\omega
^{n_{1}+n_{3}}}{n_{1}!n_{2}!n_{3}!}\text{,}\label{Iq taylor}%
\end{equation}
with the constraint $-n_{1}-\frac{1}{2}D=n_{2}+n_{3}$ coming from the
multinomial expansion.

The next step is to also expand the $q$ gaussian integral ($\ref{q momentum}$)
in Taylor series,

\begin{figure}[h]
\centering
\includegraphics[height=1.6483in,width=4.7899in]{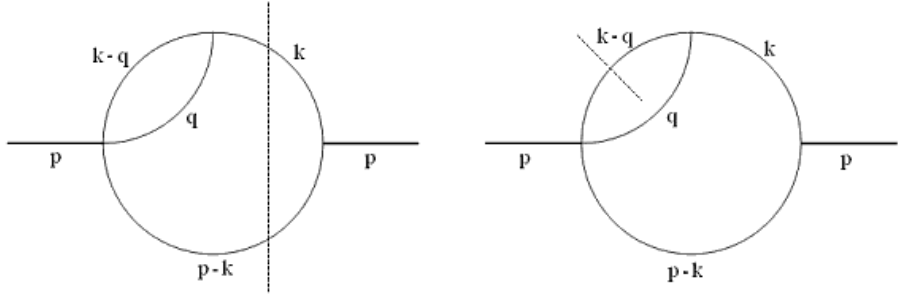}
\caption{Two-point two-loop Feynman diagrams solved using a "gluing" technique. We show two ways of cutting them in order to integrate loop-by-loop.}
\label{Figure 2:}
\end{figure}

\begin{equation}
I_{q}(k^{2};D)=\overset{\infty}{\underset{i,j=0}{\sum}}(-1)^{i+j}\frac
{\alpha^{i}\omega^{m}}{i!m!}\Im_{q}(i,m;D;k^{2})\text{,}%
\label{original taylor}%
\end{equation}
where
\begin{equation}
\Im_{q}(i,m;D;k^{2})=\int d^{D}q(q^{2})^{i}[(k-q)^{2}]^{m}\text{.}%
\label{sword fish}%
\end{equation}

This is the main sword-fish integral in negative dimension. The exponents of
the propagators are \textit{positive} and one has to consider a
\textit{negative} $D$. In the end of the calculation, one does an analytic
continuation of the result to \textit{negative} exponents and
\textit{positive} $D$. Comparing ($\ref{Iq taylor}$) and
($\ref{original taylor}$), an expression for the negative-$D$ integral can be written,%
\begin{equation}
\Im_{q}(i,m;D;k^{2})=(-\pi)^{\frac{D}{2}}\overset{\infty}{\underset{n_{i}%
=0}{\sum}}\frac{(k^{2})^{n_{1}}G}{n_{1}!n_{2}!n_{3}!}\delta_{i,n_{1}+n_{2}%
}\delta_{m,n_{1}+n_{3}}\text{,}\label{sf gama}%
\end{equation}
where a product of gamma functions is defined,%
\begin{equation}
G=\Gamma(1+i)\Gamma(1+m)\Gamma(1+i+m-2n_{1})\text{.}\label{fator gama}%
\end{equation}

The Kronecker's deltas in ($\ref{sf gama}$) and the constraint of the
multinomial expansion of ($\ref{Iq taylor}$) form a system of three equations
and three unknowns,

$\qquad\qquad\qquad\qquad\qquad\qquad\qquad\qquad\qquad\qquad\qquad
\qquad\qquad$%
\begin{equation}
\left\{
\begin{array}
[c]{l}%
n_{1}+n_{2}=i\\
n_{1}+n_{3}=m\\
n_{1}+n_{2}+n_{3}=-\frac{1}{2}D
\end{array}
\right. \label{sistema sf}%
\end{equation}
that can be easily solved with a unique solution. The integral $\Im
_{q}(i,m;D;k^{2})$ is then given by
\begin{equation}
\Im_{q}(i,m;D;k^{2})=(-\pi)^{\frac{D}{2}}(k^{2})^{i+m+\frac{1}{2}%
D}P(i,m;D)\text{,}\label{final sword fish}%
\end{equation}
where
\begin{equation}
P(i,m;D)=\frac{\Gamma(1+i)\Gamma(1+m)\Gamma(1-i-m-D)}{\Gamma(1+i+m+\frac{1}%
{2}D)\Gamma(1-i-\frac{1}{2}D)\Gamma(1-m-\frac{1}{2}D)}\text{.}\label{P sf}%
\end{equation}

With this result, it is straightforward to find the solution of the complete
flying saucer diagram. From the Taylor series ($\ref{original}$), it can be
seen that the integral in $q$ is already done, so putting
($\ref{final sword fish}$) in ($\ref{original}$), one has
\begin{equation}
\Im(i,j,l,m;D;p^{2})=(-\pi)^{\frac{D}{2}}P(i,m;D)\int d^{D}k(k^{2}%
)^{j+i+m+\frac{1}{2}D}[(p-k)^{2}]^{l}\text{,}\label{almost}%
\end{equation}
but this is exactly the integral ($\ref{sword fish}$) with the variables
$\Im_{k}(i+j+m+\frac{1}{2}D,l;D;p^{2})$, that is,
\begin{align}
\Im(i,j,l,m;D;p^{2})  & =(-\pi)^{D}(p^{2})^{i+j+l+m+D}P(i,m;D)
\nonumber\\
& \times P(i+j+m+\frac{1}{2}D,l;D)\text{,}\label{almost finished}%
\end{align}
and, finally,
\begin{align}
& \left.  \Im(i,j,l,m;D;p^{2})=(-\pi)(p^{2})^{\sigma}\frac{\Gamma
(1+i)\Gamma(1+l)\Gamma(1+m)}{\Gamma(1+\sigma)\Gamma(1-i-\frac{1}{2}%
D)\Gamma(1-m-\frac{1}{2}D)}\right. \nonumber\\
& \left.  \times\frac{\Gamma(1-\sigma-\frac{1}{2}D)\Gamma(1+i+j+m+\frac{1}%
{2}D)\Gamma(1-i-m-D)}{\Gamma(1-l-\frac{1}{2}D)\Gamma(1+i+m+\frac{1}{2}%
D)\Gamma(1+l-\sigma)}\right.  \text{,}\label{final via sf}%
\end{align}
which is exactly ($\ref{solutioncomplete}$), with $\sigma=i+j+l+m+D$. This
procedure to solve the flying saucer diagram is much simpler and faster
than solving the complete diagram as it was done in $\cite{2loops3}$.

\section{The off-shell one-loop triangle diagram}

In this section we present the other mode of cutting the complete diagram.
This is by far the most difficult and laborious way; however considering the
importance of the triangle diagram and the new result that we will show
justify the whole process. Instead of ($\ref{q momentum}$), one could begin
with the $k$ integral:
\begin{equation}
I(p^{2};D)=\int d^{D}q\exp[-\alpha q^{2}]\int d^{D}k\exp[-\beta k^{2}%
-\gamma(p-k)^{2}-\omega(k-q)^{2}]\text{,}\label{k momentum}%
\end{equation}
that corresponds to a one-loop three point function. Working on the $k$
integral, it gives without trouble,
\begin{equation}
I_{k}(p^{2},q^{2};D)=\left(  \frac{\pi}{\beta+\gamma+\omega}\right)
^{\frac{D}{2}}\exp\left[  -\frac{\beta\gamma p^{2}+\beta\omega q^{2}%
+\gamma\omega r^{2}}{\beta+\gamma+\omega}\right]  \text{,}\label{Ik}%
\end{equation}
with $r=q-p$. Expanding ($\ref{Ik}$) in Taylor series, one gets%
\begin{align}
& \left.  I_{k}(p^{2},q^{2};D)=\left(  \pi\right)  ^{\frac{D}{2}}%
\underset{n_{i}=0}{\overset{\infty}{\sum}}\frac{(-1)^{n_{1}+n_{2}+n_{3}%
}(-n_{1}-n_{2}-n_{3}-\frac{1}{2}D)!}{n_{1}!n_{2}!n_{3}!n_{4}!n_{5}!n_{6}%
!}\right. \nonumber\\
& \left.  \times(p^{2})^{n_{1}}(q^{2})^{n_{2}}(r^{2})^{n_{3}}\beta
^{n_{1}+n_{2}+n_{4}}\gamma^{n_{1}+n_{3}+n_{5}}\omega^{n_{2}+n_{3}+n_{6}%
}\right.  \text{,}\label{lk taylor series}%
\end{align}
with the constraint $n_{1}+n_{2}+n_{3}+n_{4}+n_{5}+n_{6}=-\frac{1}{2}D$ coming
from the multinomial expansion. Now one expands the original $k$ integral
($\ref{k momentum}$) in Taylor series, obtaining%
\begin{equation}
I_{k}(p^{2},q^{2};D)=\underset{j,l,m=0}{\overset{\infty}{\sum}}\frac
{(-1)^{j+l+m}}{j!l!m!}\beta^{j}\gamma^{l}\omega^{m}\Im_{k}(j,l,m;D;p^{2}%
,q^{2})\text{,}\label{triangle taylor}%
\end{equation}
where the corresponding integral in NDIM is%
\begin{equation}
\Im_{k}(j,l,m;D;p^{2},q^{2})=\int d^{D}k(k^{2})^{j}[(p-k)^{2}]^{l}%
[(k-q)^{2}]^{m}\text{.}\label{triangle}%
\end{equation}

Comparing ($\ref{lk taylor series}$) and ($\ref{triangle taylor}$) by its
$\beta$, $\gamma$ and $\omega$ powers, the integral $\Im_{k}$ has a general relation,%
\begin{align}
& \left.  \Im_{k}(j,l,m;D;p^{2},q^{2})=(\pi)^{\frac{D}{2}}(-1)^{-j-l-m}%
\overset{\infty}{\underset{n_{i}=0}{\sum}}\left[  \frac{(-1)^{n_{1}%
+n_{2}+n_{3}}G}{n_{1}!n_{2}!n_{3}!n_{4}!n_{5}!n_{6}!}\right.  \right.
\nonumber\\
& \left.  \times(p^{2})^{n_{1}}(q^{2})^{n_{2}}(r^{2})^{n_{3}}\delta
_{j,n_{1}+n_{2}+n_{4}}\delta_{l,n_{1}+n_{3}+n_{5}}\delta_{m,n_{2}+n_{3}+n_{6}%
}\right]  \text{,}\label{triangle gama}%
\end{align}
where%
\begin{equation}
G=\Gamma(1+j)\Gamma(1+l)\Gamma(1+m)\Gamma(1-n_{1}-n_{2}-n_{3}-\frac{1}%
{2}D)\text{.}\label{fator gama t}%
\end{equation}

Considering the deltas in ($\ref{triangle gama}$) and the multinomial
expansion in ($\ref{lk taylor series}$), all the constraints of the problem are%
\begin{equation}
\left\{
\begin{array}
[c]{l}%
n_{1}+n_{2}+n_{4}=j\\
n_{1}+n_{3}+n_{5}=l\\
n_{2}+n_{3}+n_{6}=m\\
n_{1}+n_{2}+n_{3}=\phi
\end{array}
\right.  \text{,}\label{sistema t}%
\end{equation}
where $\phi=j+l+m+\frac{1}{2}D$.

Therefore one has a system of four equations and six variables. There are
C$_{2}^{6}=15$ possible ways to solve the system leaving two free variables,
that is, ending up with a double series. From these $15$ possibilities, $3$
have zero determinant, so there are only $12$ non-trivial solutions that can
be grouped together according to the ratio of the momenta, or in other words,
according to the kinematical configuration. Usually, the groups are made by
four solutions and each one is linked to the others considering the symmetries
of the diagram. As said before, the leftover variables of the sum form a
double series that can be written in terms of Appel hypergeometric functions
$F_{4}$ \cite{gradstein},%
\begin{equation}
F_{4}\left(  a,b,c,d;x,y\right)  =\underset{m,n=0}{\overset{\infty}{\sum}%
}\frac{(a)_{m+n}(b)_{m+n}}{(c)_{m}(d)_{n}}\frac{x^{m}y^{n}}{m!n!}%
\text{,}\label{f4}%
\end{equation}
with the Pochhammer symbol designated by%
\begin{equation}
(a|m)\equiv(a)_{m}=\frac{\Gamma(a+m)}{\Gamma(a)}\text{,}\label{pochhammer}%
\end{equation}
and obeying the useful relations%
\begin{align}
(a|m+n)  & =(a|m)(a+m|n)\nonumber\\
& \nonumber\\
(a|-m)  & =\frac{\left(  -1\right)  ^{m}}{\left(  1-a\mid m\right)  }%
\text{.}\label{pocchammer properties}%
\end{align}

In terms of these definitions, the first set of solutions is%
\begin{align}
& \left.  \Im_{k}^{1}(j,l,m;D;p^{2},q^{2})=(-\pi)^{\frac{D}{2}}\right.
\nonumber\\
& \left.  \times\left[  A_{1}F_{4}\left(  -j,-\phi,1+l-\phi,1+m-\phi
;\frac{q^{2}}{r^{2}},\frac{p^{2}}{r^{2}}\right)  \right.  \right. \nonumber\\
& \left.  +A_{2}F_{4}\left(  -l,-j-l+\phi,1-l+\phi,1+m-\phi;\frac{q^{2}}%
{r^{2}},\frac{p^{2}}{r^{2}}\right)  \right. \nonumber\\
& \left.  +A_{3}F_{4}\left(  -m,-j-m+\phi,1+l-\phi,1-m+\phi;\frac{q^{2}}%
{r^{2}},\frac{p^{2}}{r^{2}}\right)  \right. \nonumber\\
& \left.  \left.  +A_{4}F_{4}\left(  -l-m+\phi,\phi+\frac{1}{2}D,1-l+\phi
,1-m+\phi;\frac{q^{2}}{r^{2}},\frac{p^{2}}{r^{2}}\right)  \right.  \right]
\text{,}\label{first solution}%
\end{align}
where the multiplicative factors are
\begin{equation}%
\begin{array}
[c]{l}%
A_{1}=\left(  r^{2}\right)  ^{\phi}\frac{\left(  1+\phi\mid-2\phi-\frac{1}%
{2}D\right)  }{\left(  1+l\mid-\phi\right)  \left(  1+m\mid-\phi\right)  }\\
\\
A_{2}=\left(  q^{2}\right)  ^{\phi}\left(  \frac{q^{2}}{r^{2}}\right)
^{-l}\frac{\left(  1+\phi-l\mid-2\phi+l-\frac{1}{2}D\right)  }{\left(  1+j\mid
l-\phi\right)  \left(  1+m\mid-\phi\right)  }\\
\\
A_{3}=\left(  p^{2}\right)  ^{\phi}\left(  \frac{p^{2}}{r^{2}}\right)
^{-m}\frac{\left(  1+\phi-m\mid-2\phi+m-\frac{1}{2}D\right)  }{\left(  1+j\mid
m-\phi\right)  \left(  1+l\mid-\phi\right)  }\\
\\
A_{4}=\left(  \frac{p^{2}q^{2}}{r^{2}}\right)  ^{\phi}\left(  \frac{p^{2}%
}{r^{2}}\right)  ^{-m}\left(  \frac{q^{2}}{r^{2}}\right)  ^{-l}\frac{\left(
1-j-\frac{1}{2}D\mid2j+\frac{1}{2}D\right)  }{\left(  1+l\mid j+\frac{1}%
{2}D\right)  \left(  1+m\mid j+\frac{1}{2}D\right)  }%
\end{array}
\label{factors}%
\end{equation}
This result agrees with the one calculated in the first reference in
\cite{mb1}. The other two sets with the remaining eight solutions can be found
making the replacements%
\begin{equation}%
\begin{array}
[c]{l}%
\Im_{k}^{2}(j,l,m;D;p^{2},q^{2})=\Im_{k}^{1}(q\leftrightarrow
r,j\leftrightarrow l)\\
\Im_{k}^{3}(j,l,m;D;p^{2},q^{2})=\Im_{k}^{1}(p\leftrightarrow
r,j\leftrightarrow m)
\end{array}
\label{other solutions}%
\end{equation}

Now, it comes the crucial new step that will permit us to complete the integral
($\ref{k momentum}$). As presented in ($\ref{first solution}$), the solution
is not valid in the whole space, nor takes into account the momentum
conservation $r=q-p$. This last constraint subtly reduces the number of functions $F_{4}$ from \emph{four} to \emph{three} because those original four are not linearly
independent as they should be to form a basis. Two of them can be rewritten in
terms of another function $F_{4}$ that pertains to one of the other sets of
different kinematical regions ($\ref{other solutions}$). So, with the
constraint of momentum conservation, there are only \emph{three} linearly
independent hypergeometric solutions $F_{4}$ and they hold on the whole space.
In order to reduce the four functions $F_4$ in the solution ($\ref{first solution}$), we need to combine two of them to give another $F_4$ in a different kinematical region. The right way to do this combining is given in \cite{circuito} considering the analogy between Feynman diagrams and electric circuits. Then, applying this to our case, we need to keep the first and the
third terms, and the second and fourth ones are combined together using the relation
\begin{align}
& \left.  F_{4}\left(  a,b,c,d;x,y\right)  =\frac{(b)_{-a}}{(d)_{-a}}%
(-y)^{-a}F_{4}\left(  a,a+1-d,c,a+1-b;\frac{x}{y},\frac{1}{y}\right)  +\right.
\nonumber\\
& \left.  +\frac{(a)_{-b}}{(d)_{-b}}(-y)^{-b}F_{4}\left(
b+1-d,b,c,b+1-a;\frac{x}{y},\frac{1}{y}\right)  \right. \label{relation}%
\end{align}
to see that the resulting $F_4$ is exactly a function that
appears in $\Im_{k}^{3}(j,l,m;D;p^{2},q^{2})$ given in (\ref{other solutions}). So ($\ref{first solution}$) is in fact given by
\begin{align}
& \left.  \Im_{k}^{1}(j,l,m;D;p^{2},q^{2})=(-\pi)^{\frac{D}{2}}\times\right.
\nonumber\\
& \left.  \times\left[  A_{1}F_{4}\left(  -j,-\phi,1+l-\phi,1+m-\phi
;\frac{q^{2}}{r^{2}},\frac{p^{2}}{r^{2}}\right)  +\right.  \right. \nonumber\\
& \left.  +A_{3}F_{4}\left(  -m,-j-m+\phi,1+l-\phi,1-m+\phi;\frac{q^{2}}%
{r^{2}},\frac{p^{2}}{r^{2}}\right)  +\right. \nonumber\\
& \left.  \left.  +A_{5}F_{4}\left(  -l,-m-l+\phi,1+j-\phi,1-l+\phi
;\frac{r^{2}}{p^{2}},\frac{q^{2}}{p^{2}}\right)  \right.  \right]
\text{,}\label{solution with 3}%
\end{align}
where 
$$A_{5}=\left(  q^{2}\right)  ^{\phi}\left(  \frac{q^{2}}{p^{2}}\right)
^{-l}\frac{\left(  1+\phi-l\mid-2\phi+l-\frac{1}{2}D\right)  }{\left(  1+m\mid
l-\phi\right)  \left(  1+j\mid-\phi\right)  }$$.

The kinematic region of the last term is distinct from the previous ones, but as can be seen in
Fig. 3, these two regions are connected, allowing the momentum to hold from
$-\infty$ to $+\infty$.%

\begin{figure}[h]
\centering
\includegraphics[height=3.3209in,width=3.2854in]{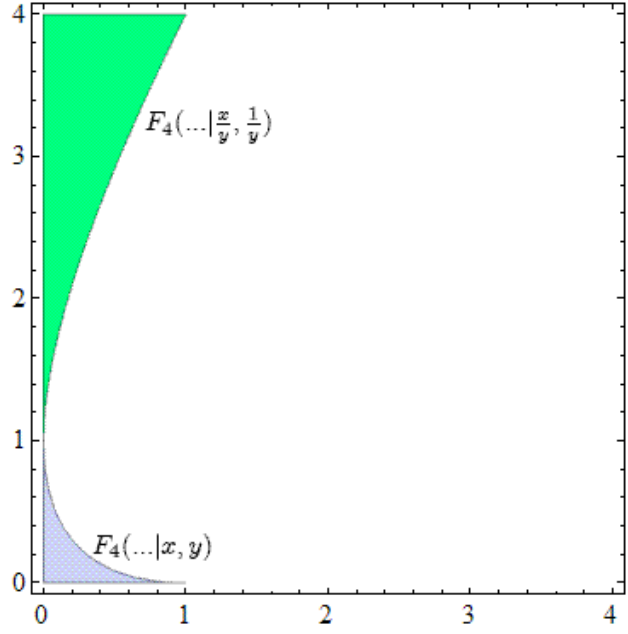}
\caption{Region of convergence of the hypergeometric series F4 when one does the correct transformation in order to account the momentum constraint.}
\label{Figure 3:}
\end{figure}

With the solution ($\ref{solution with 3}$) in hands, it is now possible to
finish the work. Going back to ($\ref{original}$), the $k$ integral is done,
so putting ($\ref{solution with 3}$) in it,%
\begin{equation}
\Im(i,j,l,m;D;p^{2})=\int d^{D}q(q^{2})^{i}\Im_{k}^{1}(j,l,m;D;p^{2}%
,q^{2})\text{,}\label{almost t}%
\end{equation}
which gives us three integrals of self-energy type, e.g. ($\ref{sword fish}$),%
\begin{align}
& \left.  \Im(i,j,l,m;D;p^{2})=(-\pi)^{\frac{D}{2}}\underset{a,b=0}%
{\overset{\infty}{\sum}}\int d^{D}q\left\{  A_{1}^{^{\prime}}(q^{2}%
)^{i+a}\left[  \left(  q-p\right)  ^{2}\right]  ^{\phi-a-b}+\right.  \right.
\nonumber\\
& \left.  \left.  +A_{3}^{^{\prime}}(q^{2})^{i+a}\left[  \left(  q-p\right)
^{2}\right]  ^{m-a-b}+A_{5}^{^{\prime}}(q^{2})^{\phi+i-l+a}\left[  \left(
q-p\right)  ^{2}\right]  ^{b}\right\}  \text{,}\right. \label{as 3 integrais}%
\end{align}
where\bigskip%
\begin{align}
A_{1}^{^{\prime}}  & =\frac{(-j)_{a+b}(-\phi)_{a+b}\left(  1+\phi\right)
_{-2\phi-\frac{1}{2}D}}{(1+l-\phi)_{a}(1+m-\phi)_{b}\left(  1+l\right)
_{-\phi}\left(  1+m\right)  _{-\phi}}\frac{\left(  p^{2}\right)  ^{b}}%
{a!b!}\nonumber\\
A_{3}^{^{\prime}}  & =\frac{(-m)_{a+b}(-j-m+\phi)_{a+b}\left(  1+\phi
-m\right)  _{-2\phi+m-\frac{1}{2}D}}{(1+l-\phi)_{a}(1-m+\phi)_{b}\left(
1+j\right)  _{m-\phi}\left(  1+l\right)  _{-\phi}}\frac{\left(  p^{2}\right)
}{a!b!}^{\phi+b-m}\nonumber\\
A_{5}^{^{\prime}}  & =\frac{(-l)_{a+b}(-m-l+\phi)_{a+b}\left(  1+\phi
-l\right)  _{-2\phi+l-\frac{1}{2}D}}{(1+j-\phi)_{a}(1-l+\phi)_{b}\left(
1+m\right)  _{l-\phi}\left(  1+j\right)  _{-\phi}}\frac{\left(  p^{2}\right)
^{l-a-b}}{a!b!}\label{coeficientes}%
\end{align}

Using the solution ($\ref{final sword fish}$) with the right replacements of
the variables, one gets%

\begin{equation}
\Im(i,j,l,m;D;p^{2})=I_{1}+I_{3}+I_{5}\text{,}\label{soma}%
\end{equation}
where%
\begin{align}
& \left.  I_{1}=(-\pi)^{D}\left(  p^{2}\right)  ^{i+\phi+\frac{1}{2}D}%
\frac{\left(  1+\phi\right)  _{-2\phi-\frac{1}{2}D}}{\left(  1+l\right)
_{-\phi}\left(  1+m\right)  _{-\phi}}\right. \nonumber\\
& \left.  \times\frac{\Gamma(1+i)\Gamma(1+\phi)\Gamma(1-i-\phi-D)}%
{\Gamma(1+i+\phi+\frac{1}{2}D)\Gamma(1-i-\frac{1}{2}D)\Gamma(1-\phi-\frac
{1}{2}D)}\right. \nonumber\\
& \left.  \times\overset{\infty}{\underset{a,b=0}{\sum}}\frac{(-j)_{a+b}%
(1+i)_{a}(i+\frac{1}{2}D)_{a}(1-i-\phi-D)_{b}(-i-\phi-\frac{1}{2}D)_{b}%
}{a!b!(1+l-\phi)_{a}(1+m-\phi)_{b}(1-\phi-\frac{1}{2}D)_{a+b}}\right.
\label{I1}%
\end{align}

\begin{align}
& \left.  I_{3}=(-\pi)^{D}\left(  p^{2}\right)  ^{i+\phi+\frac{1}{2}D}%
(-1)^{m}\frac{\left(  1+\phi\right)  _{-2\phi-\frac{1}{2}D}(-\phi)_{m}%
}{\left(  1+j\right)  _{-\phi}\left(  1+l\right)  _{-\phi}\left(
1+j-\phi\right)  _{m}}\right. \nonumber\\
& \left.  \times\frac{\Gamma(1+i)\Gamma(1+m)\Gamma(1-i-m-D)}{\Gamma
(1+i+m+\frac{1}{2}D)\Gamma(1-i-\frac{1}{2}D)\Gamma(1-m-\frac{1}{2}D)}%
\overset{\infty}{\underset{a,b=0}{\sum}}\left[  \frac{(1+i)_{a}}{a!b!}%
\right.  \right. \nonumber\\
& \left.  \left.  \times\frac{(1-i-m-D)_{b}(i+\frac{1}{2}D)_{a}(-i-m-\frac
{1}{2}D)_{b}(-j-m+\phi)_{a+b}}{(1+l-\phi)_{a}(1-m+\phi)_{b}(1-m-\frac{1}%
{2}D)_{a+b}}\right]  \right. \label{!3}%
\end{align}

\begin{align}
& \left.  I_{5}=(-\pi)^{D}\left(  p^{2}\right)  ^{i+\phi+\frac{1}{2}D}%
(-1)^{l}\frac{\left(  1+\phi\right)  _{-2\phi-\frac{1}{2}D}(-\phi)_{l}%
}{\left(  1+j\right)  _{-\phi}\left(  1+m\right)  _{-\phi}\left(
1+m-\phi\right)  _{l}}\times\right. \nonumber\\
& \left.  \times\frac{\Gamma(1)\Gamma(1+i-l+\phi)}{\Gamma(1+i-l+\phi+\frac
{1}{2}D)\Gamma(1-\frac{1}{2}D)}\overset{\infty}{\underset{a,b=0}{\sum}}\left[
\frac{(1)_{b}(1+i-l+\phi)_{a}}{a!b!(1+i-l+\phi+\frac{1}{2}D)_{a+b}}%
\times\right.  \right. \nonumber\\
& \left.  \left.  \times\frac{\left(  i-l+\phi+\frac{1}{2}D\right)  _{a}%
(\frac{1}{2}D)_{b}(-l)_{a+b}(-l-m-\phi)_{a+b}}{(1-l-+)_{a}(1+j-\phi
)_{b}(i-l+\phi+D)_{a+b}}\right]  \right.  \text{.}\label{I5}%
\end{align}

The summations above can not be written in terms of a known hypergeometric
series with defined properties. So one has to specify the values of the
indices $i,j,l,m$ and make the analytic continuation. Using
($\ref{pocchammer properties}$), it can be seen that due to the analytic
continuation of the terms $\frac{\Gamma(1)}{\Gamma(1-\frac{1}{2}D)}$ in
($\ref{I5}$), $I_{5}$ vanishes. Making $i=-1$ in $I_{1}$ and $I_{3}$, the
terms $(1+i)_{a}=\frac{\Gamma\left(  a\right)  }{\Gamma\left(  0\right)  }$
cancel the sum in the index $a$ and one has
\begin{align}
& \left.  I_{1}=(-\pi)^{D}\left(  p^{2}\right)  ^{i+\phi+\frac{1}{2}D}%
\frac{\left(  1+\phi\right)  _{-2\phi-\frac{1}{2}D}}{\left(  1+l\right)
_{-\phi}\left(  1+m\right)  _{-\phi}}\right. \nonumber\\
& \left.  \times\frac{\Gamma(1+i)\Gamma(1+\phi)\Gamma(1-i-\phi-D)}%
{\Gamma(1+i+\phi+\frac{1}{2}D)\Gamma(1-i-\frac{1}{2}D)\Gamma(1-\phi-\frac
{1}{2}D)}\right. \nonumber\\
& \left.  \times\text{ }_{2}F_{1}\left(  -j,1-i-\phi-D,1+m-\phi;1\right)
\right.  \text{,}\label{I1 i=-1}%
\end{align}

\begin{align}
& \left.  I_{3}=(-\pi)^{D}\left(  p^{2}\right)  ^{i+\phi+\frac{1}{2}D}%
(-1)^{m}\frac{\left(  1+\phi\right)  _{-2\phi-\frac{1}{2}D}(-\phi)_{m}%
}{\left(  1+j\right)  _{-\phi}\left(  1+l\right)  _{-\phi}\left(
1+j-\phi\right)  _{m}}\right. \nonumber\\
& \left.  \times\frac{\Gamma(1+i)\Gamma(1+m)\Gamma(1-i-m-D)}{\Gamma
(1+i+m+\frac{1}{2}D)\Gamma(1-i-\frac{1}{2}D)\Gamma(1-m-\frac{1}{2}D)}%
\right. \nonumber\\
& \left.  \times\text{ }_{2}F_{1}\left(  1-i-m-D,-j-m+\phi,1-m+\phi;1\right)
\right.  \text{,}\label{I3 i=-1}%
\end{align}
where%
\begin{equation}
_{2}F_{1}\left(  a,b,c;x\right)  =\underset{m=0}{\overset{\infty}{\sum}}%
\frac{(a)_{m}(b)_{m}}{(c)_{m}}\frac{x^{m}}{m!}\label{2f1}%
\end{equation}
is a hypergeometric series that can be summed up when it has unit
argument,
\begin{equation}
_{2}F_{1}\left(  a,b,c;1\right)  =\frac{\Gamma\left(  c\right)  \Gamma\left(
c-a-b\right)  }{\Gamma\left(  c-a\right)  \Gamma\left(  c-b\right)  }%
\text{.}\label{2f1 unit}%
\end{equation}

Using the property ($\ref{2f1 unit}$) in ($\ref{I1 i=-1}$, $\ref{I3 i=-1}$),
making the analytic continuation and specifying $i=j=l=m=-1$, the
integral ($\ref{soma}$) becomes
\begin{align}
& \left.  \Im_{AC}(i,j,l,m;D;p^{2})=(\pi)^{D}\left(  p^{2}\right)  ^{D-4}%
\frac{\Gamma^{3}\left(  \frac{1}{2}D-1\right)  \Gamma\left(  D-3\right)
}{\Gamma\left(  3-\frac{1}{2}D\right)  \Gamma\left(  D-2\right)  }%
\times\right. \nonumber\\
& \left.  \times\frac{\Gamma\left(  2-\frac{1}{2}D\right)  \Gamma\left(
4-D\right)  }{\Gamma\left(  \frac{3}{2}D-4\right)  }\right.  \text{,}%
\label{final}%
\end{align}
which gives the exact result of the flying saucer diagram if one makes the
analytic continuation and the specification of the exponents in the solution
($\ref{solutioncomplete}$).

\section{Conclusion}

We presented in this work a technique to integrate multi-loop Feynman
integrals. In our approach one can carry out the integrals loop-by-loop using
well-known results of straightforward one-loop diagrams. We point out the
advantage that each diagram when properly calculated --- with off-shell
external legs and respecting the constraint on the external momenta --- can be
used to solve an even harder diagram and so forth. In this approach each
diagram can be considered as a building block of another one, with more legs
and/or loops. The same method can also be applied to the two-loop master
integral and a result valid for arbitrary exponents of propagators can be obtained.

\paragraph*{Acknowledgments}

J. D. Bolzan wishes to thank CNPq (Conselho Nacional de Desenvolvimento Cient\'{\i}fico e
Tecnol\'{o}gico) for financial support. AGMS gratefully acknowledges Brazilian agencies CNPq (projects
312000/2006-5 and 471018/2007-4) and FAPERJ (projects E26/171.191-2007 and E26/170.374-2007).

\end{document}